\documentstyle[11pt]{article}

\advance\voffset by -1.5cm
\advance\hoffset by -1.5cm
\def\be{\begin{equation}}
\def\ee{\end{equation}}

\def\Zop{\bbbz}

\def\pmb#1{\setbox0=\hbox{#1}%
 \kern-.025em\copy0\kern-\wd0
 \kern.05em\copy0\kern-\wd0
 \kern-.025em\raise.0433em\box0 }
\def\A{{\cal A}}

\def\3{\ss}
\def\sq{\hbox{\rlap{$\sqcap$}$\sqcup$}}
\def\qed{\ifmmode\sq\else{\unskip\nobreak\hfil
\penalty50\hskip1em\null\nobreak\hfil\sq
\parfillskip=0pt\finalhyphendemerits=0\endgraf}\fi}
\def\half {\frac{1}{2}}
\def\quart {\frac{1}{4}}

\def\bbbz {{\sf Z\!\!Z}}

\newcommand{\ket}[1]{|#1\rangle}
\newcommand{\bra}[1]{\langle#1|}
\def\Tr{{\rm Tr}}

\begin{document}

\thispagestyle{empty}
\def\thefootnote{\fnsymbol{footnote}}
\begin{flushright}
  hep-th/9701137\\
  HUTP-97/A003 \\
  BRX TH-402\\
\end{flushright} \vskip 1.0cm

\begin{center}\LARGE
{\bf A Non-Supersymmetric Open String Theory and S-Duality}
\end{center} \vskip 1.0cm
\begin{center}\large
       Oren Bergman%
       $^{a,b}$\footnote{E-mail  address: {\tt bergman@string.harvard.edu}}
       and 
       Matthias R.\ Gaberdiel%
       $^{b}$\footnote{E-mail  address: {\tt gaberd@string.harvard.edu}}
\end{center}
\vskip0.5cm

\begin{center}\it$^a$
Department of Physics \\
Brandeis University \\
Waltham, MA 02254
\end{center}
\begin{center}\it$^b$
Lyman Laboratory of Physics\\
Harvard University\\
Cambridge, MA 02138
\end{center}
\vskip 1em
\begin{center}
January 1997
\end{center}
\vskip 1cm
\begin{abstract}
A non-supersymmetric ten-dimensional open string 
theory is constructed
as an orbifold of type I string theory, and as an
orientifold of the bosonic type B theory.
It is purely bosonic, and cancellation of massless tadpoles
requires the gauge group to be $SO(32)\times SO(32)$. The spectrum
of the theory contains a closed string tachyon, and open string
tachyons in the $({\bf 32},{\bf 32})$ multiplet. 
The $D$-branes of this theory are analyzed, 
and it is found that the massless
excitations of one of the 1-branes coincide with the world-sheet 
degrees of freedom of the $D=26$ bosonic string theory compactified
on the $SO(32)$ lattice. This  suggests that the two theories are
related by S-duality.
\end{abstract}

\vfill

\setcounter{footnote}{0}
\def\thefootnote{\arabic{footnote}}

\newpage

\renewcommand{\theequation}{\thesection.\arabic{equation}}

\section{Introduction}
\setcounter{equation}{0}

The past two and a half years have seen a tremendous increase in our
understanding of the dynamics of superstring
theory \cite{Rev}. In particular it has become apparent that the
five ten-dimensional theories, together with an eleven-dimensional 
theory (M-theory), are different limits in moduli space of some unifying
description.  A guiding principle throughout has been the fact that
supersymmetry guarantees that certain perturbative results are
protected from corrections, and are therefore exact. Another
important insight has been the realization that certain solitonic
states admit a fully stringy description in terms of Dirichlet
branes \cite{PCJ}. 
Similar advances have not occurred for non-super\-sym\-met\-ric
theories, as they are not known to possess
non-renormalization theorems. On
the other hand, some of them have $D$-branes, and therefore should
allow for an analogous analysis of their non-perturbative dynamics.

In ten dimensions, there are a number
of closed non-super\-sym\-met\-ric string
theories which can be obtained as certain orbifolds
of the four closed supersymmetric theories.  These include 
the bosonic type A and type B theories,
which are orbifolds by $(-1)^{F_S}$ of type IIA and type IIB,
where $F_S$ is the spacetime fermion number
operator \cite{SeiWit,DixHar}. There are furthermore seven
non-supersymmetric heterotic string theories, which are obtained as
orbifolds by $(-1)^{F_S} \cdot y$ from the two heterotic theories,
where $y$ is an automorphism of the weight lattice \cite{DixHar} (see
also \cite{AGMV,KLT,FHPV}); these theories are uniquely characterized
by their gauge group, and the different possible groups are $SO(32)$,
$E_8\times SO(16)$, $SO(24)\times SO(8)$, $(E_7 \times SU(2))^2$,
$SU(16)\times U(1)$, $E_8$ and $SO(16)\times SO(16)$. All of these
theories, except for the last one, have tachyons.  By analyzing 
modular properties it has been shown that
these are the only consistent non-supersymmetric
closed string theories in ten dimensions
which can be obtained from a free fermionic construction \cite{KLT}.

It is well known that there is also an open supersymmetric
theory in ten dimensions, the so-called type I theory with
$N=1$ supersymmetry and gauge group $SO(32)$. It can be obtained 
as an orientifold of type IIB \cite{SagCar,Horava,Dai,Pol95}. It is 
therefore natural to ask whether there exists an analogous
non-supersymmetric open theory which may be obtained
as an orientifold from type B. Furthermore, in view
of the diagram,
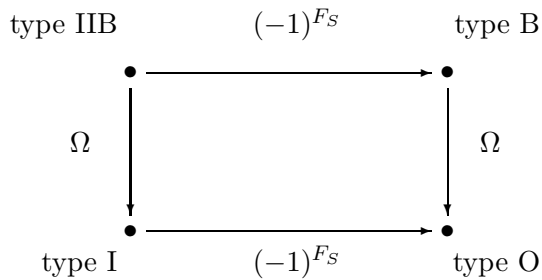
\begin{displaymath}
\begin{picture}(175,150)(-90,-40)
\put(-60,0){\vbox to 0pt
        {\vss\hbox to 0pt{\hss$\bullet$\hss}\vss}}
\put(60,0){\vbox to 0pt
        {\vss\hbox to 0pt{\hss$\bullet$\hss}\vss}}
\put(-60,60){\vbox to 0pt
        {\vss\hbox to 0pt{\hss$\bullet$\hss}\vss}}
\put(60,60){\vbox to 0pt
        {\vss\hbox to 0pt{\hss$\bullet$\hss}\vss}}
\put(-60,54){\vector(0,-1){48}}
\put(60,54){\vector(0,-1){48}}
\put(-54,60){\vector(1,0){108}}
\put(-54,0){\vector(1,0){108}}
\put(-65,-15){\hbox to 0pt{\hss$\mbox{type I}$}}
\put(95,-15){\hbox to 0pt{\hss$\mbox{type O}$}}
\put(-65,75){\hbox to 0pt{\hss$\mbox{type IIB}$}}
\put(95,75){\hbox to 0pt{\hss$\mbox{type B}$}}
\put(-75,30){\hbox to 0pt{\hss$\Omega$}}
\put(80,30){\hbox to 0pt{\hss$\Omega$}}
\put(20,-15){\hbox to 0pt{\hss$(-1)^{F_S}$}}
\put(20,75){\hbox to 0pt{\hss$(-1)^{F_S}$}}
\end{picture}
\end{displaymath}
the theory in question should also be an orbifold of the
type I theory by $(-1)^{F_S}$. It is one of the aims
of this paper to exhibit these two constructions, 
thereby describing this open string theory, which
we shall call type O, and its $D$-branes in detail. This theory
was first constructed by Bianchi and Sagnotti \cite{Sag}
from the orientifold point of view.\footnote{We thank 
Carlo Angelantonj for drawing our attention to this paper.}
It will turn out that tadpole cancellation conditions require
the gauge group of type O to be $SO(32)\times SO(32)$. 
Since type O theory is non-chiral, and actually free of fermions,
the low-energy effective theory is
trivially anomaly free.  
\smallskip

One of the best established strong/weak coupling dualities of 
supersymmetric theories is the duality of the
heterotic $SO(32)$ closed string theory with
the type I theory \cite{Wit95a,PolWit}. Both
theories have the same low-energy effective
action, and in particular the same gauge group $SO(32)$.
The strongest
piece of evidence for this duality is the fact that
the type I theory possesses a 1-brane, whose massless
excitations coincide with the world-sheet degrees of
freedom of the heterotic string \cite{PolWit}. 

Type O has gauge group $SO(32)\times SO(32)$,
and if its strong coupling dual is yet another string theory, 
the analogy to the previous case suggests that this
string theory is closed and that it
has gauge group $SO(32)\times SO(32)$.
The only candidate is the $D=26$ bosonic string theory compactified
to ten dimensions on the $SO(32)$ lattice. 
Analyzing the branes of type O, we find that the theory
has four different 1-branes, and 
that the massless excitations of one of them indeed reproduce
the world-sheet excitations of the bosonic 
string theory in 26 dimensions compactified on the $SO(32)$ 
weight lattice. (The degrees of freedom of this 1-brane 
can be thought of as corresponding to roughly double
the excitations of the heterotic string.) 
Furthermore, the two theories
have identical tachyonic spectra, and parts of the
massless spectra agree. On the other hand, the $D=26$ bosonic
theory has additional scalars, and type O has an additional
massless $2$-form, but as we are considering
non-supersymmetric theories, there is a priori no reason
why the low-energy effective degrees of freedom should match up.

We regard this as evidence for the proposal that the strong
coupling limit of the type O theory is the 26 dimensional bosonic
theory compactified on the $SO(32)$ weight lattice. To our
knowledge this is the first evidence for a duality between
non-supersymmetric theories, although the $D=26$ bosonic
string, together with the superstrings
and the supermembrane, has already made an appearance as a possible 
target of the $(2,1)$ heterotic string \cite{KuMar}.
If true, this duality would indicate
that these non-supersymmetric string theories might be 
non-perturbatively consistent despite having tachyons. 
It would also suggest that string duality does not necessarily
rely on supersymmetry.
\smallskip

The paper is organized as follows. In section 2 we explain how the
type O theory can be obtained as the orbifold of type I. We analyze
the tadpole condition carefully, and explain why the gauge group is
$SO(32)\times SO(32)$. In section 3 we demonstrate that the theory
can also be obtained as an orientifold of type B. In particular, we
show that the open strings necessary to cancel the tadpoles can be
obtained from boundary states of the type B theory which are invariant
under world-sheet parity. In section 4 we analyze the massless
excitations of the 1-branes of this theory, and show that for one of
them they coincide with the world-sheet degrees of freedom of the
bosonic string in 26 dimensions compactified on the $SO(32)$ lattice.
We also analyze the tachyonic and massless degrees of freedom of both
theories. In section~5 we make some conclusive remarks, and 
we have included three appendices, where we give details
of the various $D$-brane calculations.

\section{Type O as a Type I Orbifold}
\setcounter{equation}{0}

\noindent
The simplest way to realize type O theory is as an
orbifold of type I string theory. Some of the features of the 
resulting type O string theory, such as the appropriate open string
GSO projection, are easy to see from this
point of view. Other features, such as the precise $D$-brane 
spectrum, are easier to analyze from the orientifold
point of view discussed in the next section.
Let us thus begin by reviewing type I open string theory. 

\subsection{Review of Type I string theory}
\noindent
Recall that type I string theory consists of unoriented
open strings with GSO projection $\half (1+(-1)^F)$ in both the
Ramond (R) and Neveu-Schwarz (NS) sectors, and unoriented closed
strings with GSO projection $\quart (1+(-1)^F)(1+(-1)^{\widetilde{F}})$
in the R-R, NS-NS, R-NS and NS-R sectors, resulting in
an $N=1$ supersymmetric theory.
The physical states of the unoriented strings
must also be invariant under the action of the world-sheet parity 
operator $\Omega$, which on the closed string sector is defined by
\cite{GimPol}
\be
\label{Omegaonclosed}
\begin{array}{lcl}
{\displaystyle \Omega \alpha_{n}^{\mu} \Omega}
 & = & {\displaystyle \widetilde{\alpha}_{n}^{\mu}} \\
{\displaystyle \Omega \psi_{r}^{\mu} \Omega} & = &
{\displaystyle \widetilde{\psi}_{r}^{\mu}}  \\
{\displaystyle \Omega \widetilde{\psi}_r^\mu \Omega}
 &= & {\displaystyle -\psi_r^\mu \,,}
\end{array}
\ee
so that
\begin{equation}
\label{Omega}
\Omega \psi_{r}^{\mu} \widetilde{\psi}_{r}^{\nu}\Omega  =
   \psi_{r}^{\nu} \widetilde{\psi}_{r}^{\mu}\,,
\end{equation}
and 
\begin{eqnarray}
\label{Omegaonclosedgs}
  \Omega |0\rangle_{NS}\otimes |0\rangle_{NS} &= &
  |0\rangle_{NS}\otimes |0\rangle_{NS} \nonumber \\
  \Omega |S^\alpha\rangle_{R}\otimes |\widetilde{S}^\beta\rangle_{R} &=& 
  - |S^\beta\rangle_{R}\otimes |\widetilde{S}^\alpha\rangle_{R}\,,
\end{eqnarray}
where $\ket{S^\alpha}_R$ and 
$\ket{\widetilde{S}^\beta}_R$ are states in the 
spinor representation ${\bf 8}_s \oplus{\bf 8}_c$. 
These conventions are 
chosen so that in the NS-NS (R-R) sector the symmetric (anti-symmetric)
combinations are invariant under $\Omega$. 
Thus the massless 
closed string sector consists of the metric $G_{\mu\nu}$, dilaton
$\Phi$, R-R 2-form $A_{\mu\nu}$, and their super-partners.

The massless sector of the unoriented open string 
consists of the $SO(N)$ gauge field 
$A_\mu^a$ and its super-partner. Cancellation of the
massless tadpoles requires $N=32$, and thus the gauge group
to be $SO(32)$ \cite{GS,PolCai,CLNY}. This can also be understood
as an anomaly cancellation condition in the low-energy effective
action \cite{GrSchw}.

Type I string theory contains Dirichlet $p$-branes for $p=1,5,9$. 
This follows, for example, 
from the fact that type I can be understood as the orientifold 
of type IIB string theory, together with the transformation properties
of the various boundary states in the type IIB theory
(see the analogous analysis for type O in section~3.2). 
{}From this point of view,
tadpole cancellation requires the inclusion of 
$32$ 9-branes in the vacuum which give rise to the
appropriate Chan-Paton factors for the open strings.

\subsection{Type O string theory}
Now let us gauge the $\Zop_2$ symmetry group 
of the type I theory corresponding to a $2\pi$ rotation in space.
Because of the spin-statistics theorem, this transformation is 
equivalently described by $(-1)^{F_S}$.
The effect of the corresponding orbifold 
in string theory is to remove all the
fermions from the spectrum, namely the R sector of the open string
and the R-NS and NS-R sectors of the closed string,
thus completely breaking supersymmetry. The resulting
untwisted sector is given by unoriented open strings in the NS sector and
unoriented closed strings in the R-R and NS-NS sectors, with the same GSO 
projections (and Chan-Paton factors) as before.
For the closed strings we have to introduce a twisted sector
in the standard way, where again the orbifold 
projection removes all fermions. This twisted sector
is most easily described in the Green-Schwarz formulation 
following \cite{DixHar}. It corresponds to (unoriented)
strings which close only up to a transformation by $(-1)^{F_S}$.
This means that the spacetime spinors $S$ and 
${\widetilde S}$ are both
anti-periodic. The lowest state in this sector is a tachyon,
and the next states are given by the massless states
\begin{equation}
 S^\alpha_{-1/2}\ket{0}\otimes {\widetilde S}^\beta_{-1/2}\ket{0} \,.
\end{equation}
Here $S^\alpha_{-1/2}\ket{0}$ and 
${\widetilde S}^\beta_{-1/2}\ket{0}$ 
are states in the same spinor representation, which is
of opposite chirality to the physical spinor representation in the
untwisted sector; in the RNS formalism, these states
therefore correspond to states in the NS-NS and R-R sectors,
where we choose the opposite GSO projection from above, namely
$\quart (1-(-1)^F)(1-(-1)^{\widetilde{F}})$. 

The spectrum of the orbifolded theory is thus purely bosonic, and
consists of unoriented open strings in the NS sector with GSO
projection $\half (1+(-1)^F)$, and unoriented closed strings in the
R-R and NS-NS sectors with GSO projection $\half
(1+(-1)^{F+\widetilde{F}})$. Only states invariant under $\Omega$ are
physical, so in particular the massless fields are given in the open
string sector by vectors $A_\mu^a$ in the adjoint representation of
$SO(32)$, and in the closed sector by the metric $G_{\mu\nu}$, dilaton
$\Phi$, and two R-R 2-forms $A^1_{\mu\nu},A^2_{\mu\nu}$ from the
anti-symmetric combinations of ${\bf 8}_s\otimes{\bf 8}_s$ and 
${\bf 8}_c\otimes{\bf 8}_c$.  In addition there is a closed string 
tachyon with $\alpha_O' M^2 = -2$.

To analyze whether this is a consistent string theory, we have to
check whether the massless tadpoles vanish. Recall that
anomalies in the
low-energy effective theory are not an issue since there are no
chiral fields. 
However, as pointed out in \cite{PolCai}, there are more
fundamental reasons to cancel tadpoles; in particular the R-R tadpole
must vanish to satisfy the equation of motion of the ten-form gauge
field. The NS-NS (dilaton) tadpole is less severe, as it can,
in principle, be removed by a shift of the background (Fischler-Susskind 
mechanism) \cite{FS}, but this introduces a spacetime-dependent 
coupling constant; in the following we shall therefore attempt to
cancel both the NS-NS and R-R massless tadpoles. Unlike the situation in
type I string theory, where the total vacuum amplitude vanishes
because of supersymmetry, this will not happen in our case. 
As a consequence, we have to analyze the tadpoles in the
R-R and NS-NS sectors separately.

The relevant contributions to the one-loop vacuum amplitude are given by
\be
 \begin{array}{rcl}
  A_C &=& {\displaystyle
      \int_0^\infty {dt\over 2t}\Tr_{open}\left[
      e^{-2\pi tL_0}(-1)^{F_S}P^{GSO}_{open}{1\over 2}\right]} \\[10pt]
  A_M &=& {\displaystyle
      \int_0^\infty {dt\over 2t}\Tr_{open}\left[
      e^{-2\pi tL_0}(-1)^{F_S}P^{GSO}_{open}{\Omega\over 2}\right]}
      \\[10pt]
  A_K &=& {\displaystyle
      \int_0^\infty {dt\over 2t}\Tr_{closed}\left[
      e^{-2\pi t(L_0 + \widetilde{L}_0)}
      (-1)^{F_S}P^{GSO}_{closed}{\Omega\over 2}
      \right]\,, }
 \end{array}
\ee
corresponding to the cylinder, M\"obius strip, and Klein bottle,
respectively. In order to extract the massless tadpoles 
we must perform a modular transformation,
relating the loop channel calculation to a tree channel 
calculation involving the exchange of closed strings between
either boundary or crosscap external states.

It is useful to summarize the effect of the modular transformation
in a ``translation table'', relating the contributions
in the tree channel to the corresponding contributions in the loop
channel. This correspondence depends on which sectors are
actually present in the theory, as well as on the 
GSO projection in each sector. As a consequence, the translation 
table is different for type O theory, where it is
given by
\begin{center}
\begin{tabular}{|l||l|l|} 
\hline
Type O & tree channel & loop channel \\ \hline\hline
Cylinder & $R\otimes R$ & $NS;\, \half (-1)^F\times\half$ \\ 
\mbox{}  & $NS\otimes NS$ & $NS; \, {1\over 4}$ \\ \hline
M\"obius strip & $R\otimes R$ & 0 \\
\mbox{}  & $NS\otimes NS$ & $NS; \, \half (1+(-1)^F)
          \times\half\Omega$ \\ \hline
Klein bottle & $R\otimes R$ & 0 \\ 
\mbox{}  & $NS\otimes NS$ & $(R\otimes R + NS\otimes NS); \,
                      \half(1+(-1)^{F+\widetilde{F}})
                      \times\half\Omega\,,$ \\
\hline
\end{tabular}
\end{center}
compared with the corresponding table for type I theory \cite{PolCai}
\begin{center}
\begin{tabular}{|l||l|l|} 
\hline
Type I  & tree channel & loop channel \\ \hline\hline
Cylinder & $R\otimes R$ & $NS;\, \half (-1)^F\times\half$ \\ 
\mbox{}  & $NS\otimes NS$ & $(NS - R);\, {1\over 4}$ \\ \hline
M\"obius strip & $R\otimes R$ & $ R; \,  -\half\times\half\Omega $\\
\mbox{}  & $NS\otimes NS$ & $NS; \, \half (1+(-1)^F)
          \times\half\Omega$ \\ \hline
Klein bottle & $R\otimes R$ & $(R\otimes R + NS\otimes NS); \,
                      {1\over 4}((-1)^F+(-1)^{\widetilde{F}})
                      \times\half\Omega$ \\ 
\mbox{}  & $NS\otimes NS$ & $(R\otimes R + NS\otimes NS); \,
                      {1\over 4}(1+(-1)^{F+\widetilde{F}})
                      \times\half\Omega\,.$ \\
\hline
\end{tabular}
\end{center}
The different factors for the Klein bottle 
amplitudes for type O theory compared to type I theory 
deserve some comment. In the type I theory, the GSO projection 
in the closed string sector is 
$\quart (1 + (-1)^F + (-1)^{\widetilde{F}} + (-1)^{F+\widetilde{F}})$; 
the NS-NS tree channel contribution corresponds to the 
first and the last term in the loop channel trace,
and the R-R tree channel contribution corresponds to the second 
and the third term in the loop channel trace. On the other hand, for the 
type O theory the GSO projection in the closed string sector is just
$\half (1+(-1)^{F+\widetilde{F}})$, and therefore there is no R-R tree
contribution, and the NS-NS tree contribution is twice as large.

In the type O case, only the cylinder has a R-R contribution, and it
is therefore clear that it must cancel by itself. 
The cylinder corresponds to the tree channel amplitude
involving two boundary 9-branes. As we
have already $32$ 9-branes in the theory (which now carry
only bosonic open string excitations), the only way to cancel 
the R-R tadpole is to
introduce an equal number of anti-9-branes, as branes and anti-branes
carry opposite R-R charge. This means, in particular, that the
relevant gauge group will be a product group $SO(N)\times SO(N)$,
where $N\geq 32$.  (As will become apparent from the discussion of the
next section, the low-lying excitations of open stings stretching
between branes and branes, and anti-branes and anti-branes are
massless vectors, whereas open strings stretching between branes and
anti-branes have an open string tachyon, and no massless excitations.
As we are considering an unoriented theory, the vectors give rise
to an $SO(N)$ group.)
In order to decide whether there is a solution for $N$, we have to
analyze the various contributions to the massless NS-NS tadpole.

{}From the table above we see that the NS-NS vacuum amplitudes are 
given by
\be
\begin{array}{rcl}
 A^{NSNS}_C &=& {\displaystyle
         {1\over 4}\int_0^\infty{dt\over 2t}
         \Tr_{NS}\Big[e^{-2\pi tL_0}\Big]} \\[10pt]
 A^{NSNS}_M &=& {\displaystyle
         {1\over 4}\int_0^\infty{dt\over 2t}
         \Tr_{NS}\Big[e^{-2\pi tL_0}(1+(-1)^F)\Omega\Big]}
         \\[10pt]
 A^{NSNS}_K &=& {\displaystyle
         {1\over 4}\int_0^\infty{dt\over 2t}
         \Tr_{NSNS+RR}\Big[e^{-2\pi t(L_0+\widetilde{L}_0)}
        (1+(-1)^{F+\widetilde{F}})\Omega\Big]\,.}
\end{array}
\ee
The cylinder trace is a straightforward computation
which is similar to the
type I case except for the absence of the R sector open strings
\begin{equation}
 A^{NSNS}_C = {G_C\over 4}\int_0^\infty{dt\over 2t}
     (8\pi^2t)^{-5}{f_3(e^{-\pi t})^8\over f_1(e^{-\pi t})^8}\,.
\end{equation}
Here $G_C$ is the group factor for the cylinder, which for the
gauge group $SO(N)\times SO(N)$ is $G_C= (2 N)^2$, and the functions 
$f_i$ and their modular properties are given in eqs.~(C.1,C.2) 
of appendix~C. 
The M\"obius strip trace is exactly the same as in the type I case
 \begin{equation}
 A^{NSNS}_M = {G_M\over 4}\int_0^\infty{dt\over 2t}
     (8\pi^2t)^{-5}{f_2(ie^{-\pi t})^8\over f_1(ie^{-\pi t})^8}\,,
\end{equation}
where $G_M$ is the group factor for the M\"obius strip, which in
our case is $G_M=-2N$.
The Klein bottle trace is twice what it is in type I; the action
of $\Omega$ in the closed string trace 
(\ref{Omegaonclosed},\ref{Omegaonclosedgs})
entails that only states which are left-right symmetric 
contribute, for which $(-1)^{F+\widetilde{F}}$ is automatically $+1$.
Furthermore, because of the relative minus
sign in the action of $\Omega$ on the R-R and NS-NS
ground states, the two sectors appear with opposite sign,
and the resulting expression has the structure of the
difference of an open string NS trace and an open string
R trace, both without any $(-1)^F$ insertions. Using the 
abstruse identity (\ref{abstruse}), we thus find
\begin{equation}
 A^{NSNS}_K = {1\over 2}\int_0^\infty{dt\over 2t} (4\pi^2t)^{-5}
           {f_4(e^{-2\pi t})^8\over f_1(e^{-2\pi t})^8} \,,
\end{equation}
and there is no group factor.  

To extract the singular behavior at $t=0$ we substitute in the
integrals
\begin{equation}
l = \left\{
    \begin{array}{ll}
      1/2t & {\mbox{cylinder}} \\
      1/8t & {\mbox{M\"obius strip}} \\
      1/4t & {\mbox{Klein bottle}\,,}
    \end{array}
    \right.
\end{equation}
and perform the appropriate modular transformation, thus
obtaining the tree channel expression for the NS-NS vacuum amplitude
\begin{equation}
A^{NSNS} = -(8\pi^2)^{-5}\int{dl\over 2}\left[
   {G_C\over 2}{f_3(e^{-2\pi l})^8\over f_1(e^{-2\pi l})^8} +
   32G_M {f_2(ie^{-2\pi l})^8\over f_1(ie^{-2\pi l})^8} +
   32^2 {f_2(e^{-2\pi l})^8\over f_1(e^{-2\pi l})^8} \right]\,.
\end{equation}
The $l\rightarrow\infty$ behavior of the modular functions is manifest
from their definition (\ref{fi}). The relevant
terms in the expression in brackets are then 
\begin{equation}
 \half G_C e^{2\pi l} + 4(G_C + 4\cdot 32G_M + 4\cdot 32^2) 
= 2 N^2 e^{2\pi l} + 16 (32 - N)^2\,,
\end{equation}
where we have inserted the actual values of the group
theory factors. The first term is the tachyon tadpole, which we 
expect in a theory with a closed string tachyon, and 
the other terms correspond to the massless NS-NS tadpole that 
we want to cancel. We can see directly that the massless
tadpole cancels for $N=32$.

Type O string theory is thus a theory of unoriented open and closed
strings with a gauge group $SO(32)\times SO(32)$. Its massless fields
consist of a vector transforming in the adjoint representation
$A_\mu^a$, a metric $G_{\mu\nu}$, a dilaton $\Phi$, and a pair of
2-forms $A^{1,2}_{\mu\nu}$.  The ground state in the closed string
sector is  a singlet tachyon with $\alpha_O' M^2 = -2$.  In addition,
open strings stretching between 9-branes and anti-9-branes give rise
to a  $({\bf 32},{\bf 32})$ multiplet of tachyons with 
$\alpha_O' M^2 = -1/2$. 

The allowed $D$-branes appear to be the same as in type I, namely
$p=1,5,9$, except that now they do not carry fermionic modes. We
will see in the following section that this is not quite correct, and 
that the $D$-brane spectrum of type O string theory is somewhat 
richer. A first indication of this is the fact that type O
has {\it two} 2-forms in the R-R sector. This will be 
important for the 1-brane analysis in section 4.

\section{Type O as an Orientifold of Type B}
\setcounter{equation}{0}

\noindent
In this section we want to discuss how type O theory can
also be obtained as an orientifold of type B theory.
The construction is in close analogy to the realization
of type I as an orientifold of type IIB.
As in the previous construction, massless tadpoles will have to
be cancelled, and this will require the inclusion of
$32$ 9-branes and $32$ anti-9-branes. The actual calculation
will be identical to the one we have performed in the previous
section, and we shall therefore not repeat it. The point
of the analysis of this section is to demonstrate that there 
exist appropriate branes leading to the correct open string
sector in the spectrum of type O theory. 
Along the way, we shall also, more generally, discuss the
different Dirichlet branes of type O and type B theory.

\subsection{Type B string theory and its $D$-branes}
\noindent
Type B string theory is obtained from type IIB just as type O was from
type I, namely by a $(-1)^{F_S}$ orbifold \cite{DixHar}. It is a
purely bosonic theory of oriented closed strings in the R-R and NS-NS
sectors with GSO projection $\half (1+(-1)^{F+\widetilde{F}})$. The NS-NS
sector contains a tachyon with $\alpha' M^2 = -2$.  The massless fields
consist of a metric $G_{\mu\nu}$, Kalb-Ramond field $B_{\mu\nu}$, and
dilaton $\Phi$ in the NS-NS sector, and two scalars $A^{1,2}$, two
2-forms $A^{1,2}_{\mu\nu}$, and a self-dual and an
anti-self-dual 4-form $A^{+}_{\mu\nu\lambda\rho}$, 
$A^{-}_{\mu\nu\lambda\rho}$ in the R-R sector. 

The R-R sector of type B is effectively doubled compared
to the R-R sector of type IIB, and in the NS-NS sector there
is an additional tachyon. We therefore expect that there exist
additional boundary states (compared to the boundary states
of type IIB) which are charged under the additional R-R fields,
and which couple to the tachyon (rather than to the dilaton
and the graviton). As we are dealing with a non-supersymmetric theory, 
the NS-NS and R-R sectors are in essence independent, and we expect 
that the theory will have four boundary $p$-branes for each odd
$p$ (as well as four anti-$p$-branes).\footnote{The corresponding
statement also holds for the type A theory for even $p$.}

The explicit construction and analysis of the boundary states
can be found in appendix~A. A general boundary
$p$-brane state is of the form
\be 
\alpha_{+} \ket{Bp,+}_{NSNS} 
+ \alpha_{-} \ket{Bp,-}_{NSNS}
+ \beta_{+} \ket{Bp,+}_{RR} 
+ \beta_{-} \ket{Bp,-}_{RR}\,,
\label{boundarystates} 
\ee 
where $p$ is odd, and $\alpha_{\pm}$ and $\beta_{\pm}$ are 
arbitrary normalization constants. (For type IIB, the
normalization constants are fixed up to an overall
scale, as explained in appendix~A.)
However, physical $D$-brane states also have to satisfy
another consistency condition.\footnote{We thank 
Joe Polchinski for drawing our attention to this problem.
(See \cite{Pol96} for related issues.) 
We also thank Cumrun Vafa and Barton Zwiebach
for helpful discussions on this point.}
The additional condition comes from the
fact that open strings which end on the same boundary state should 
be able to close, and that
the resulting state should already be in the closed-string spectrum
of the theory. For the type B theory, this implies, in particular,
that the corresponding open string states should only be in the NS
sector, as the open R sector excitations are fermions which are
absent in the type B theory. We also expect, from the analogy
of the situation for type IIB and type I, that the GSO projection
of the NS open string sector should agree with the GSO projection
in the type O theory. 

The implications of these two constraints can be easily read off
from the analysis of appendix~C. The open string sector
corresponding to strings beginning and ending on the same boundary state
is determined by the loop channel corresponding to the tree channel
amplitude of the boundary state with itself. 
The first constraint implies then 
that the NS-NS sector contribution is either $\ket{Bp,+}_{NSNS}$ or
$\ket{Bp,-}_{NSNS}$, but not a linear combination of them. Likewise,
the second constraint determines the correct relative normalization
of the NS-NS and the R-R boundary states in (\ref{boundarystates})
up to a phase. The actual linear combination of 
$\ket{Bp,+}_{RR}$ and $\ket{Bp,-}_{RR}$ 
for the R-R boundary state is not fixed by this consideration, as
\be
\bra{Bp,+}e^{-lH}\ket{Bp,-}_{RR} = 0 \,,
\ee
and so $\ket{Bp,+}_{RR}$ and $\ket{Bp,-}_{RR}$ do not mix. We can
therefore, without loss of generality, restrict ourselves to
considering R-R boundary states which are either 
$\ket{Bp,+}_{RR}$ or $\ket{Bp,-}_{RR}$. We should mention
that the two different R-R boundary states for given $p$
are charged with respect to the two different $(p+1)$-forms in 
the R-R sector.

This leaves us with a total of {\it four} Dirichlet branes for 
each odd $p$ in type B string theory, given by
\be
\ket{Bp,\eta}_{NSNS} + \ket{Bp,\eta^\prime}_{RR}\,,
\label{BDbranes}
\ee 
where $\eta$ and $\eta^\prime$ are independently $+$ or $-$. 
We denote these by $|Bp,\eta\eta'\rangle$. 
For anti-$D$-branes the sign in front of the R-R boundary state 
is reversed.

\subsection{Type O string theory and its $D$-branes}
\noindent 
In taking the orientifold projection of type B string theory we remove
the states which are not invariant under $\Omega$, and add 9-branes as
necessary to cancel the massless tadpoles arising from the
non-orientable world-sheets. Given the $\Omega$ transformation
conventions of (\ref{Omegaonclosed}) and (\ref{Omegaonclosedgs}), the
first step leaves us with a massless sector consisting of a metric
$G_{\mu\nu}$, dilaton $\Phi$, and two 2-forms $A^{1,2}_{\mu\nu}$,
and the closed-string tachyon of $\alpha_O' M^2=-2$. The
massless tadpole calculation is the same as in the previous section,
and requires the addition of 32 9-branes and 32 anti-9-branes for
cancellation. 

The spectrum of $D$-branes in type O string theory consists
of those boundary states of the form (\ref{BDbranes}) which
are invariant under $\Omega$.  
In the NS-NS sector it follows from the explicit expression
for the boundary state (\ref{solution}) together with 
eqs.~(2.1-2.3) that
$$
\Omega |Bp,\eta\rangle_{NSNS} = |Bp,\eta\rangle_{NSNS}\,.
$$
To understand the action in the R-R sector, we have to
analyze carefully how $\Omega$ acts on the 
boundary ground state 
$|B7,+\rangle_{RR}^0$ which is defined in appendix~A;
this is explained in appendix~B. We then find that
\be
\begin{array}{lcl}
\Omega |Bp,\eta\rangle^0_{RR} & = & {\displaystyle
\Omega \prod_{\mu=p+3}^{9} (\psi_0^{\mu}
+ i \eta \widetilde{\psi}_0^{\mu})  |B7,\eta\rangle^0_{RR} } \\
& = & {\displaystyle - \prod_{\mu=p+3}^{9}
(-i\eta)(\psi_0^{\mu} + i \eta \widetilde{\psi}_0^{\mu})
|B7,\eta\rangle^0_{RR}} \\
& = & {\displaystyle - (-i\eta)^{7-p} |Bp,\eta\rangle^0_{RR} \,,}
\end{array}
\ee
and it follows that 
\be
\Omega |Bp,\eta\rangle_{RR} = - (-i \eta)^{7-p} |Bp,\eta\rangle_{RR} \,.
\ee
Thus only Dirichlet $p$-branes with $p=1,5$ and $9$ survive the
orientifold projection. (The same analysis also applies to the
orientifold of type IIB theory, thus establishing that type I
theory has $p$-branes for $p=1,5,9$.) For the type O theory, 
there are four branes of each such $p$ as explained in the 
previous subsection. 

The theory therefore contains 9-branes and anti-9-branes,
and we are able to cancel the massless tadpoles by the 
same calculation as in section~2. Indeed, as explained before,
there are four different Dirichlet 9-branes, and we can cancel
the massless tadpoles by including 32 9-branes of a given
type, as long as the anti-9-branes are of the {\it same}
type. To see that this will reproduce the calculation
of section~2, we note that the open strings stretching
between two 9-branes (\ref{BDbranes}) of the same type
are in the GSO-projected NS sector (with
the GSO projection being $\half (1 +(-1)^F)$), whereas
the open strings stretching between a 9-brane 
and an anti-9-brane of the same type are in the NS sector with 
projection $\half (1 - (-1)^F)$, as follows from the calculations of
appendix~C. This matches up with the assumed structure of the
open string sector in the calculation of section~2.

This argument also shows that the open strings stretching
between branes and branes or anti-branes and anti-branes
give rise to massless vector
fields in the adjoint representation of $SO(32)\times SO(32)$.
On the other hand, the open strings stretching between
branes and anti-branes do not have massless excitations,
but give rise to tachyons of $\alpha_O' M^2=-1/2$ in the
$({\bf 32},{\bf 32})$ multiplet of $SO(32)\times SO(32)$.
This completes the description of the spectrum of the
type O theory.

{}From the point of view of the orientifold, there exist
other solutions which cancel the massless tadpoles
(but which cannot be obtained as orbifolds of type I). We
can consider $(32-n)$ 9-branes of type $|B9,++\rangle$
and $n$ 9-branes of type $|B9,--\rangle$, together with
their respective anti-branes. The resulting open string
theory then possesses the gauge group 
$SO(32-n)\times SO(n)\times SO(32-n)\times SO(n)$, 
open string tachyons in the 
$({\bf 32-n},{\bf 1},{\bf 32-n},{\bf 1})$ and 
$({\bf 1}, {\bf n}, {\bf 1}, {\bf n})$
representations, and massless fermions in the 
$({\bf 32-n}, {\bf n}, {\bf 1}, {\bf 1})$,
$({\bf 32-n}, {\bf 1}, {\bf 1}, {\bf n})$,
$({\bf 1}, {\bf n}, {\bf 32-n}, {\bf 1})$
and $({\bf 1}, {\bf 1}, {\bf 32-n}, {\bf n})$ representations
of the gauge group, as well as the same closed string spectrum
as type O.
The fermions are excitations of open strings stretching between branes
or anti-branes of type $|B9,++\rangle$ and branes or anti-branes
of type $|B9,--\rangle$; in particular these open strings cannot close, 
and there
is no inconsistency with the fact that the closed string sector does
not contain fermions.
These theories were first constructed
by Bianchi and Sagnotti \cite{Sag}.

\section{A proposal for the strong coupling limit}
\setcounter{equation}{0}

In  this section we want to analyze the massless excitations of the
1-branes of type O string theory. These are described as in
\cite{PolWit} by two different types of open strings: those beginning
and ending on the 1-brane, and those beginning on the 1-brane and
ending on the
background 9-branes. As regards the former, all different 1-branes are
alike; it follows from the analysis in appendix~C ({\it cf.}
(\ref{c.10},\ref{NSNSpp-a},\ref{RRpp+},\ref{RRpp-})) 
that the open string corresponding to the tree diagram
$$
\langle B1,\eta\eta'| e^{-lH} |B1,\eta\eta'\rangle
$$
is in the NS sector with the GSO projection $\half (1+(-1)^F)$; 
there are thus
no tachyonic modes, and eight massless bosonic modes in ${\bf 8}_v$.
By the same argument as in \cite{PolWit}, there are then
eight left-moving and eight right-moving scalars on
the world-sheet of the 1-brane.

On the other hand, the open string sector relating the 1-branes to the
background 9-branes depends on the relation between the 1-brane and
the background 9-branes. In defining the type O theory, we have to
make a choice, by taking the 9-branes (and anti-branes) to be of a
definite type. To fix notation for the following, let us assume,
without loss of generality, that
these 9-branes are all of type $|B9,++\rangle$.

For the 1-branes of type $|B1,++\rangle$ there are no additional 
massless modes, as the contribution from the R-R boundary state
vanishes (\ref{RR19+}), and as the contribution from the
NS-NS boundary state is purely massive (\ref{NSNS19+a}).
The massless excitations of the $|B1,++\rangle$
1-brane are then eight left- and right-moving scalars, and
therefore do not constitute the world-sheet degrees of freedom
of a critical string theory.

For the 1-branes of type $\ket{B1,-+}$ there are additional
massless modes in the twisted R sector
(where all eight transverse bosons
and fermions are half-integrally moded),
coming from the NS-NS boundary state
(\ref{NSNS19-a}). Each twisted R sector has two massless states,
which, in a Lorentz-covariant formulation, transform as a spinor
of the two longitudinal directions. The two states have opposite chirality,
and therefore correspond to a left-moving fermion and a right-moving fermion
on the world-sheet of the 1-brane. As there are $32$ $9$-branes and
$32$ anti-$9$-branes, we end up with a total of $64$ left-moving
and $64$ right-moving fermions, 
as well as the eight left- and right-moving scalars. Again,
this does not correspond to a critical string theory.

For $\ket{B1,+-}$, the situation is similar to the previous case, 
except that the additional massless modes come from the R-R boundary 
state, and are in the twisted R sector with $(-1)^F$ (\ref{RR19-}).
This gives the same number of massless fermions as in the previous
case, although the significance of the 
various minus signs in the trace is unclear.
At any rate, this does not give a critical string theory.

This leaves us with the $|B1,--\rangle$ 1-brane, for which 
we have a contribution to
the massless states both from the R-R sector, and the NS-NS
sector (\ref{NSNS19-a},\ref{RR19-}). 
Altogether, we get $32$ open strings in the twisted R sector
with $\half (1+(-1)^F)$ (from the $32$ 9-branes), and $32$ open 
strings in the twisted R sector with $\half (1-(-1)^F)$ (from the 
$32$ anti-9-branes). 
Each of these sectors has a single massless fermion of
definite chirality, the $9$-branes giving rise to fermions of
opposite chirality to those coming from the anti-$9$-branes.
We therefore obtain in addition to the eight left- and
right-moving scalars, $32$ left- and $32$ right-moving fermions
in the $({\bf 32},{\bf 1})$ and $({\bf 1},{\bf 32})$ of
$SO(32)\times SO(32)$, respectively. 
These
are precisely the world-sheet excitations of the $26$-dimensional
bosonic string theory compactified on the $SO(32)$ lattice.

As we do not have supersymmetry at our disposal to extend tree-level
calculations to strong coupling, we do not know how the tension of the
different 1-branes behave at strong coupling. However, if we are
willing to assume that the strong coupling regime of the type O
theory is again described by another string theory, the above
analysis seems to suggest that at strong coupling the $|B1,--\rangle$
1-brane becomes light, and that the strong coupling limit of the
type O theory is described by the $26$ dimensional bosonic theory
compactified on the $SO(32)$ lattice.\footnote{We should mention that
the $|B1,--\rangle$ 1-brane seems to be the only 1-brane for which the 
massless excitations correspond to any known string theory, even if 
we relax the open-closed consistency condition and allow arbitrary
linear combinations of boundary states.}

A similar analysis can also be performed for the 
$(SO(32-n)\times SO(n))^2$ theories, which were 
discussed at the end of section~3. In this case, none of the
$1$-branes have massless excitations which correspond
to the world-sheet degrees of freedom of any (known) 
critical string theory. This is reassuring, as these
theories have massless fermions in addition to a rank
$32$ gauge group.

As discussed in sections 2 and 3, type O theory has one singlet
tachyon of mass squared $\alpha_O' M^2=-2$, and a $({\bf 32}, {\bf 32})$
multiplet of tachyons of mass squared $\alpha_O' M^2=-1/2$. On the
other hand, we can describe the $26$ dimensional
bosonic theory compactified on the $SO(32)$ lattice
in terms of eight (left-moving) 
bosonic oscillators $\alpha_n^{\mu}$, where
$\mu=2, \ldots, 9$ and $n\in\Zop$, and  $32$ (left-moving)
fermionic oscillators $\psi^a_{r}$, where $a=1, \ldots 32$, 
$r\in \Zop + 1/2$ in the NS sector and $r\in\Zop$ in
the (purely massive) R sector, 
together with their corresponding right-movers. It is
then easy to see that the theory has one (singlet)
tachyon of mass squared $\alpha_B' M^2 = -4$ (the ground state
$|0\rangle\otimes |0\rangle$), and a $({\bf 32}, {\bf 32})$
multiplet of tachyons of mass squared $\alpha_B' M^2 = -2$,
corresponding to the states $\psi^a_{-1/2} \widetilde{\psi}^b_{-1/2}
|0\rangle\otimes |0\rangle$. The tachyons of the two theories
therefore appear in the same representations.

As regards massless states, both theories have a metric, Kalb-Ramond
field and dilaton, and the same gauge fields. (For the $26$ dimensional
theory, these are the states of the form 
$\alpha_{-1}^{\mu} \widetilde{\alpha}_{-1}^{\nu} 
|0\rangle \otimes |0\rangle$ and 
$\psi^a_{-1/2} \psi^b_{-1/2} \widetilde{\alpha}_{-1}^{\mu}
|0\rangle \otimes |0\rangle$, 
$\alpha_{-1}^{\mu} \widetilde{\psi}^a_{-1/2} \widetilde{\psi}^b_{-1/2} 
|0\rangle \otimes |0\rangle$.) In particular,
the $D=26$ bosonic theory is the only known closed string theory 
which can have a perturbative gauge group of rank $32$.

Apart from these (rather surprising) coincidences, there
are also differences in the massless spectrum.\footnote{This 
does not rule out the proposed
strong coupling relation, as the theory in question is not 
supersymmetric; in particular, there is no reason
why the masses of the states should not get renormalized.}
In particular, the $26$ dimensional 
theory has massless scalars in the bi-adjoint representation 
of $SO(32)\times SO(32)$, that are not present in type O; 
these are the states of the form
$$\psi^a_{-1/2} \psi^b_{-1/2} \widetilde{\psi}^c_{-1/2}
\widetilde{\psi}^d_{-1/2}|0\rangle\otimes|0\rangle\,.$$ 
On the other hand, type O theory has an additional
2-form in the R-R sector, which couples to the two 
1-branes $|B1,\pm+\rangle$. 
As regards
the former, these scalars are the moduli of the Narain lattice
describing the compactification. As has been pointed out by Ginsparg
and Vafa \cite{GinVafa}, the cosmological constant is critical for
the points in the moduli space of Narain lattices for which the
theory has maximal gauge group (such as $SO(32)\times SO(32)$), and it
is therefore conceivable that the theory is frozen at such a
point for large coupling. This would imply that these scalars become
massive at large coupling. As regards the additional 2-form,
the situation is rather less clear.

\section{Conclusions}
\setcounter{equation}{0}

In this paper we have analyzed an open string theory in ten
dimensions which can be obtained as an
orbifold of type I, and as an orientifold of type B, which
itself is an orbifold of type IIB. This theory is a special case
of a class of orientifolds which were first described by 
Bianchi and Sagnotti \cite{Sag}. The open string theory,
type O, has a closed string tachyon of $\alpha_O' M^2=-2$ and open
string tachyons of $\alpha_O' M^2=-1/2$ in the $({\bf 32},{\bf 32})$
multiplet of the gauge group $SO(32)\times SO(32)$. Apart from this
gauge group, the other massless states are the dilaton, graviton, and
two 2-forms. 

We have analyzed the $D$-branes of type B and of type O.
These theories have essentially the same $D$-brane spectrum as
type IIB and type I, respectively, apart from the fact that
for every (allowed) $p$, type B and type O possess four distinct
Dirichlet $p$-branes instead of just one. We have explained that
there exists an additional consistency condition
related to the compatibility of open and closed strings, which singles
out certain boundary states as the physical $D$-branes.

The relation of the four theories, type IIB, B, I and O,
to each other has a natural interpretation in terms of F-theory
\cite{Vafa}. By definition, compactification of F-theory on a 
torus is type IIB, and it has been argued in \cite{MorVaf} that
compactification of F-theory on a cylinder is type I.
We can obtain the cylinder as a $\Zop_2$-quotient of the torus,
where the $\Zop_2$ group acts on the coordinates of the
torus $(x,y), x\sim x+R_x, y \sim y+ R_y$, as 
$i_x: (x,y) \mapsto (-x,y)$. This is consistent with the
fact that type I can be obtained as the $\Zop_2$
orientifold of type IIB.

There exist two quotient manifolds of the torus which do not
have covariantly constant spinors, and it is natural
to assume that compactification of F-theory on these will
lead to non-supersymmetric string theories in ten dimensions.
One of them is the (closed) tetrahedron, which is the $\Zop_2$
quotient of the torus by the action $(x,y)\mapsto (-x,-y)$,
and the other is the (open) rectangle, which is
the $\Zop_2\times\Zop_2$ quotient of the torus by the
action of $i_x$ and $i_y$, where $i_y$ is defined by
$(x,y)\mapsto (x,-y)$. Following the above reasoning,
it is suggestive to identify the first theory with
type B, the $\Zop_2$ orbifold of type IIB, and 
the second with type O, the $\Zop_2$ orientifold
of type B or the $\Zop_2$ orbifold of type I. This
picture is consistent as we can obtain the rectangle
as a $\Zop_2$ quotient of the cylinder (by the action
of $i_y$) and of the tetrahedron (by the action
of $i_x$). 

{}From this point of view, it is suggestive to associate the gauge
group $SO(32)$ to a {\it pair} of opposite boundaries. The closed
theories, type IIB and B, do not have any gauge fields, the cylinder
theory (type I) has one pair of boundaries and gauge group $SO(32)$,
and the rectangle theory (type O) has two pairs of boundaries and
gauge group $SO(32)\times SO(32)$. Obviously, this discussion is
rather formal, and much more work is needed to substantiate these
suggestions.
\medskip

We analyzed the massless excitations of the
four 1-branes of type O, and found that
for one of them, the excitations agree precisely with the
world-sheet excitations of the bosonic string in $D=26$
dimensions compactified on the $SO(32)$ lattice. Furthermore,
this $D=26$ theory and type O have the same gauge group, the
same tachyonic representations, and parts of the massless spectrum agree. 
We regard this as evidence for the proposal that the strong 
coupling limit of type O theory is this $D=26$ bosonic theory. 

We find it surprising that the tachyonic states
match. On the other hand, we do not yet  have
a good understanding of  the
mismatch of the massless scalars and 2-forms, or
the behavior of the 1-branes of type O at strong coupling.

\appendix
\section*{Appendix}

We want to explain a novel way of doing $D$-brane calculations. 
In this approach the boundary states are defined as coherent
states in the closed string theory (as in \cite{PolCai,CLNY}),
and the physical $D$-brane states are then identified by
certain consistency conditions, rather than by their
supersymmetry transformation properties. 
We shall recover many of the
known results for $D$-branes of the type IIA/IIB/I theories, 
but we can also apply these techniques to 
the non-supersymmetric type A/B/O theories.

\section{The coherent state approach to $D$-branes}
\renewcommand{\theequation}{A.\arabic{equation}}
\setcounter{equation}{0}

In the following we shall work in the RNS formalism in light-cone
gauge. This will make the formulae more manageable as there will be no
ghost contributions. On the other hand, this approach limits us to
considering only $D$-branes whose dimensions differ at most by
eight. Furthermore, the two light-cone directions must have Dirichlet
boundary conditions, so that we are really describing
$(p+1)$-instantons rather than Dirichlet $p$-branes \cite{GrGut}.
Since all boundary states have at least two Dirichlet directions, we
will be considering $p=-1, \ldots, 7$.
These instantons are related by a double Wick
rotation to $D$-branes, and we shall therefore 
assume that the results of
our calculations apply equally to the corresponding
Dirichlet branes \cite{GrGut}. In the following
we shall always refer to the boundary states as $D$-branes.

A boundary state corresponding to a Dirichlet 
$p$-brane $|Bp,\eta\rangle$ 
satisfies
\be
\label{Nboundary}
\left.
\begin{array}{lcl}
{\displaystyle 
(\alpha_n^\mu - \widetilde{\alpha}_{-n}^{\mu}) |Bp,\eta\rangle} & = & 0 \\
{\displaystyle
(\psi_r^{\mu} - i \eta \widetilde{\psi}_{-r}^{\mu}) |Bp,\eta\rangle}
& = & 0 
\end{array}
\right\} \mu=2, \ldots, p+2\,, 
\ee
\be
\label{Dboundary}
\left.
\begin{array}{lcl}
{\displaystyle
(\alpha_n^\mu + \widetilde{\alpha}_{-n}^{\mu}) |Bp,\eta\rangle} & = & 0 \\
{\displaystyle
(\psi_r^{\mu} + i \eta \widetilde{\psi}_{-r}^{\mu}) |Bp,\eta\rangle}
& = & 0
\end{array}
\right\} \mu=p+3, \ldots, 9\,,
\ee
where we have chosen $\mu=0,1$ as the light-cone directions.
Here $\eta=\pm 1$ (the two different signs correspond to different
spin structures), the $\psi_r^{\mu}$ are the
half-integrally (integrally) moded fermionic oscillators of the
left-moving NS(R) sector, and the tilde denotes the corresponding
right-moving mode. 

The solution to these equations is given by the coherent state
$$
|Bp,\eta\rangle = {\cal N}\exp\left\{  \sum_{n=1}^{\infty}
\left[-{1 \over n} \sum_{\mu=2}^{p+2} \alpha_{-n}^{\mu}
\widetilde{\alpha}_{-n}^{\mu}  + {1 \over n} \sum_{\mu=p+3}^{9} 
\alpha_{-n}^{\mu} \widetilde{\alpha}_{-n}^{\mu} \right] \right.
\hspace*{4.5cm}
$$
\be
\label{solution}
\hspace*{4cm} \left.
+ i \eta \sum_{r>0}^{\infty}
\left[- \sum_{\mu=2}^{p+2} \psi_{-r}^{\mu}
\widetilde{\psi}_{-r}^{\mu}  + \sum_{\mu=p+3}^{9}
\psi_{-r}^{\mu} \widetilde{\psi}_{-r}^{\mu} \right]\right\}
|0\rangle \otimes |0\rangle\,,
\ee
where ${\cal N}$ is a normalization constant
which, for later convenience, we choose to be ${\cal N}=1$ in
the NS-NS sector and ${\cal N}=4i$ in the R-R sector, and
$|0\rangle \otimes |0\rangle$ is the ground state of 
either the NS-NS
or the R-R sector. In the latter case, this ground state 
will be denoted by $|Bp,\eta\rangle_{RR}^0$; it satisfies
(\ref{Nboundary}) and (\ref{Dboundary}) with $r=0$.

The boundary states in the NS-NS sector transform under 
$(-1)^{F}$ and $(-1)^{\widetilde{F}}$ as
\be
\label{FerNS}
\begin{array}{lcl}
(-1)^{F} |Bp,\eta\rangle_{NSNS} & = &
- |Bp,-\eta\rangle_{NSNS} \\
(-1)^{\widetilde{F}} |Bp,\eta\rangle_{NSNS} & = &
- |Bp,-\eta\rangle_{NSNS} \,,
\end{array}
\ee
as $(-1)^F |0\rangle_{NS} = - |0\rangle_{NS}$. To analyze
the situation in the R-R sector, we introduce the modes
$$\psi_{\pm}^{\mu} = {1 \over \sqrt{2}} 
\left(\psi_0^{\mu} \pm i \widetilde{\psi}_0^{\mu}\right)\,,$$
which satisfy the anti-commutation relations 
$\{\psi_{\pm}^\mu, \psi_{\pm}^\nu\}=0$
and $\{\psi_{+}^\mu, \psi_{-}^\nu\}= \delta^{\mu \nu}$, as
$\{\psi_{0}^\mu, \psi_{0}^\nu\}=
\{\widetilde{\psi}_{0}^\mu, \widetilde{\psi}_{0}^\nu\}=\delta^{\mu \nu}$
and $\{\psi_{0}^\mu, \widetilde{\psi}_{0}^\nu\}=0$. 
The chirality operators in light-cone gauge are 
$\Gamma_9=\psi_0^2 \cdots \psi_0^9$ and 
$\widetilde{\Gamma}_9=\widetilde{\psi}_0^2 \cdots \widetilde{\psi}_0^9$, and
they satisfy
$$
\begin{array}{lcllcl}
\Gamma_9 \psi_{\pm}^{\mu} & = & - \psi_{\mp}^{\mu} \Gamma_9
 \hspace*{1cm}
& \Gamma_9^2 & = & 1 \vspace*{0.3cm} \\ 
\widetilde{\Gamma}_9 \psi_{\pm}^{\mu} & = & 
\psi_{\mp}^{\mu} \widetilde{\Gamma}_9 \hspace*{1cm}
& \widetilde{\Gamma}_9^2 & = & 1\,. 
\end{array} 
$$
{}From (\ref{Nboundary})  we see that the state $|B7,+\rangle_{RR}^0$ 
satisfies
the equation $\psi^\mu_{-} |B7,+\rangle_{RR}^0 = 0$ for all
$\mu=2, \ldots, 9$. We fix the normalization of $|B7,-\rangle_{RR}^0$
by defining 
$|B7,-\rangle_{RR}^0=\widetilde{\Gamma}_9 |B7,+\rangle_{RR}^0$. It
then follows that 
$$
\begin{array}{lclcllcl}
\Gamma_9 |B7,+\rangle_{RR}^0 & = & 
i^8 \widetilde{\Gamma}_9 |B7,+\rangle_{RR}^0 &
= & |B7,-\rangle_{RR}^0 \qquad & 
\widetilde{\Gamma}_9 |B7,+\rangle_{RR}^0 & 
= & |B7,-\rangle_{RR}^0 \\[6pt]
\Gamma_9 |B7,-\rangle_{RR}^0 & = & |B7,+\rangle_{RR}^0 & & &
\widetilde{\Gamma}_9 |B7,-\rangle_{RR}^0 &
= & |B7,+\rangle_{RR}^0 \,.
\end{array}
$$
Next we define 
$$
|Bp,\pm\rangle_{RR}^0 = \prod_{\mu=p+3}^{9} \psi_{\pm}^{\mu} 
|B7,\pm\rangle_{RR}^0 \,,$$
and then $\psi_{\mp}^{\mu} 
|Bp,\pm\rangle_{RR}^0=0$ for $\mu=2, \ldots, p+2$,
and $\psi_{\pm}^{\mu} 
|Bp,\pm\rangle_{RR}^0=0$ for $\mu=p+3, \ldots, 9$. 
Furthermore, 
$$
\begin{array}{lcl}
\widetilde{\Gamma}_9 |Bp,\pm\rangle_{RR}^0 & = 
&  |Bp,\mp\rangle_{RR}^0 \\[6pt]
\Gamma_9 |Bp,\pm\rangle_{RR}^0 & = 
&  (-1)^{7-p} |Bp,\mp\rangle_{RR}^0 \,.
\end{array}
$$
This implies that the action of $(-1)^{F} \Gamma_9$ and
$(-1)^{\widetilde{F}} \widetilde{\Gamma}_9$ on the boundary state 
is given by
\be
\label{FerR}
\begin{array}{lcl}
(-1)^{F} \Gamma_9 |Bp,\eta\rangle_{RR} & = &
(-1)^{7-p} |Bp,-\eta\rangle_{RR} \\[6pt]
(-1)^{\widetilde{F}} \widetilde{\Gamma}_9 
|Bp,\eta\rangle_{RR} & = &
|Bp,-\eta\rangle_{RR} \,.
\end{array}
\ee
Taking (\ref{FerNS}) and (\ref{FerR}) together, we can now deduce
which boundary states are compatible with the GSO projection.  In the
type II theories, physical states in the NS-NS (R-R) sector are
invariant under $(-1)^{F}$ ($(-1)^{F} \Gamma_9$) and
$(-1)^{\widetilde{F}}$ ($(-1)^{\widetilde{F}} \widetilde{\Gamma}_9$)
separately; in the NS-NS sector, the only physical state is
\be
|Bp\rangle_{NSNS} = {1 \over 2}\Bigl( |Bp,+\rangle_{NSNS} -
|Bp,-\rangle_{NSNS}\Bigr) \,,
\ee
which is physical for all $p$, whereas in the R-R sector, the 
only potentially physical state is
\be
|Bp\rangle_{RR}  = {1 \over 2}\Bigl( |Bp,+\rangle_{RR} +
|Bp,-\rangle_{RR}\Bigr) \,,
\ee
which is physical for even $p$ in the case of type IIA, and
odd $p$ in the case of type IIB. 
By a similar argument to the one following eq.~(3.1), the open-closed
consistency condition requires the physical $D$-branes to be
sums of NS-NS and R-R boundary states. This then implies 
that the $D$-branes
of type IIA (type IIB) only exist for even (odd) $p$. This 
restriction can
also be understood from the fact that these $D$-branes preserve 
$1/2$ of the spacetime supersymmetry \cite{Li}.

In the bosonic type A/B theories, the states only have to
be invariant under $(-1)^{F+\widetilde{F}}$ 
($(-1)^{F+\widetilde{F}} \Gamma_9 \widetilde{\Gamma}_9$). It then follows
that in both the NS-NS and the R-R sectors
the states $|Bp,+\rangle$ and $|Bp,-\rangle$ are already
GSO-invariant, and thus form allowed boundary states. In each sector 
there are therefore two independent boundary states; this
reflects the fact that the massless fields of
the R-R sector are doubled compared to the corresponding
type II theories, and that the NS-NS sector contains in addition 
a tachyon.

\section{Orientifolded $D$-branes}
\renewcommand{\theequation}{B.\arabic{equation}}
\setcounter{equation}{0}

In this short appendix we want to determine how
$\Omega$ acts on the ground state of the $D$-brane
coherent state $|B7,+\rangle_{RR}^0$ in the R-R sector. 
(The action on the other ground states follows then
easily, as explained in section~3.2.)
Let us first introduce a convenient description for the ground
states of the R-R sector. 

Given the oscillators $\psi_0^{\mu}, \widetilde{\psi}_0^{\nu}$,
where $\mu, \nu=2, \ldots, 9$, satisfying the anti-commutation
relations as described in appendix~A, we define
\be
\begin{array}{lcll}
b^j_{\pm} & = & {1 \over \sqrt{2}} \left(
\psi_0^{2j} \pm i \psi_0^{2j+1} \right) & j=1,2,3,4 \\[6pt]
\widetilde{b}^{j}_{\pm} & = & 
{1 \over \sqrt{2}} \left(
\widetilde{\psi}_0^{2j} \pm i 
\widetilde{\psi}_0^{2j+1} \right) & j=1,2,3,4\,,
\end{array}
\ee
which satisfy the anti-commutation relations 
$\{b^i_*,\widetilde{b}^j_*\}=0$, 
$\{b^i_{\pm},b^j_{\pm}\} =
\{\widetilde{b}^i_{\pm},\widetilde{b}^j_{\pm}\} = 0$,
$\{b^i_{+}, b^j_{-} \}= \{\widetilde{b}^i_{+},\widetilde{b}^j_{-}\} 
= \delta^{ij}$. 
We can then consider the space of states created by the action
of $b^i_+$ and $\widetilde{b}^j_{+}$ from a vacuum state $|0\rangle$
which satisfies $b^i_{-} |0\rangle = \widetilde{b}^j_{-} |0\rangle=0$.
The R-R ground state space of the type B theory is then the subspace 
of this space for which the difference between the number of
$b^i_+$ and $\widetilde{b}^j_{+}$ generators is even. 

Next we introduce the generators
\be 
B^{i, \pm} = \left(b^i_{+} \pm i \widetilde{b}^i_{+} \right) 
\hspace*{1cm}
C^{i, \pm} = \left(b^i_{-} \pm i \widetilde{b}^i_{-} \right)\,.
\ee
The generators $B^{i, \pm}$ anti-commute among
themselves, and we can equivalently describe the R-R ground state
space of the type B theory as being generated from $|0\rangle$ 
by the action of an even number of $B^{i, \pm}$. (The generators
$C^{i,\pm}$ all annihilate $|0\rangle$.) 

In terms of these generators, the boundary ground state
$|B7,+\rangle_{RR}^0$ satisfies the equations
$$ B^{i,-} |B7,+\rangle_{RR}^0 = C^{i,-} |B7,+\rangle_{RR}^0 = 0\,,$$
where $i=1,2,3,4$. The unique solution for 
$|B7,+\rangle_{RR}^0$ is therefore (up to normalization)
\be
|B7,+\rangle_{RR}^0 = \prod_{i=1}^{4} B^{i,-} |0\rangle \,, 
\ee
as $C^{i,-}$ anti-commutes with $B^{j,-}$, and annihilates
$|0\rangle$. 

We can now easily read off how $\Omega$ acts on this state. First
of all, it is clear that $\Omega |0\rangle = - |0\rangle$, as
$|0\rangle$ is a left-right symmetric state and we have
chosen the convention (\ref{Omegaonclosedgs}). (This convention,
together with (\ref{Bis}) below, implies that the anti-symmetric
combination of the R-R ground states are invariant under $\Omega$.)
Furthermore, because of (\ref{Omegaonclosed}),
\be
\label{Bis}
\Omega B^{i,-} \Omega  =  i B^{i,-}\,,
\ee
and thus $\Omega |B7,+\rangle_{RR}^0 = - |B7,+\rangle_{RR}^0$.

\section{Loop and tree calculations}
\renewcommand{\theequation}{C.\arabic{equation}}
\setcounter{equation}{0}

In this appendix we collect the relevant formulae
for relating the loop- and tree-channel calculations
involving the boundary states.
Let us start by introducing some
notation which will prove convenient for the following.
Following Polchinski and Cai \cite{PolCai} (see also
\cite{PCJ}), we define
\be
\label{fi}
\begin{array}{lcl}
f_1(q) & = & {\displaystyle
q^{1/12} \prod_{n=1}^{\infty} ( 1- q^{2n}) } \\
f_2(q) & = & {\displaystyle
\sqrt{2} q^{1/12} \prod_{n=1}^{\infty} (1 + q^{2n}) } \\
f_3(q) & = & {\displaystyle
q^{-1/24} \prod_{n=1}^{\infty} (1 + q^{2n-1}) } \\
f_4(q) & = & {\displaystyle 
q^{-1/24} \prod_{n=1}^{\infty} (1 - q^{2n-1})\,.}
\end{array}
\ee
For $q=e^{-\pi t}$, the limit
$t\rightarrow \infty$ corresponds to $q\rightarrow 0$, for
which the asymptotic expansion of these functions is obvious.
The modular transformation which we shall use repeatedly in
the following is $q \mapsto \widetilde{q}=e^{-\pi/t}$, and
the relevant relations are
\be
f_1(\widetilde{q}) = \sqrt{t} f_1(q) \hspace{1cm}
f_2(\widetilde{q}) = f_4(q) \hspace{1cm}
f_3(\widetilde{q}) = f_3(q) \hspace{1cm}
f_4(\widetilde{q}) = f_2(q)\,. 
\ee
(We should mention that these identities only hold up to
a root of unity whose eighth power is $+1$.) The functions
also satisfy the so-called ``abstruse identity''
%corrected the abstruse identity
\be
\label{abstruse}
f_3(q)^8  - f_2(q)^8 - f_4(q)^8 = 0 \,.
\ee
The relevant amplitudes are then
\be
\A(p,\eta;q,\eta') = {1 \over 2} \int_{0}^{\infty} {dl \over 2}
\langle Bp,\eta | e^{-l H} | Bq, \eta' \rangle \,,
\ee
where $H=- 2 p^+ p^- + (p^\perp)^2 + M^2$, $M^2=\sum n N_n$ and
$N_n$ is the number operator for the oscillators.
Using the explicit form for the various boundary states, the
inner products can be determined; they take the form of a 
trace in an open string theory, reflecting the fact that
the boundary state relates left- and right-movers. In more
detail, we find that in the NS-NS sector
\be
\label{NSNSamp}
\langle Bp, \eta | e^{-l H} | Bq, \eta'\rangle_{NSNS} = 
Tr_{NS} \left(e^{-2 l L_0} (\eta \eta')^{F} \prod_{\mu=p+3}^{q+2}
(-1)^{B_{\mu}} \right) \,,
\ee
where we have assumed (without loss of generality) that $p\leq q$,
and where $(-1)^{B_{\mu}}$ is defined by
\be
\begin{array}{lcl}
(-1)^{B_{\mu}} \alpha^{\nu}_{m} (-1)^{B_{\mu}} & = & \left\{
\begin{array}{ll}
\alpha^{\nu}_{m} & \mu\neq \nu \\
- \alpha^{\nu}_{m} & \mu=\nu 
\end{array} \right. \\[10pt]
(-1)^{B_{\mu}} \psi^{\nu}_{r} (-1)^{B_{\mu}} & = & \left\{
\begin{array}{ll}
\psi^{\nu}_{r} & \mu\neq \nu \\
- \psi^{\nu}_{r} & \mu=\nu \,.
\end{array} \right.
\end{array}
\ee
Here $\alpha^{\nu}_{m}$ and $\psi^{\nu}_{r}$ are the bosonic
and fermionic modes in the open string trace. We shall mainly
be interested in the case of $p=q$, for which the above simplifies
to 
\be
\label{NSNSpp}
\langle Bp, \eta | e^{-l H} | Bp, \eta\rangle_{NSNS} =
{f_3(e^{-2\pi l})^8 \over f_1(e^{-2\pi l})^8}\,, \hspace*{0.3cm}
\langle Bp, \eta | e^{-l H} | Bp, -\eta\rangle_{NSNS} =
{f_4(e^{-2\pi l})^8 \over f_1(e^{-2\pi l})^8}\,.
\ee
The other case of interest is $q=p+8$, which in our case
corresponds to $p=-1$ and $q=7$. In the main body
of the paper we mainly use the following results for $p=1, q=9$; 
this can be deduced from the results for $p=-1, q=7$ 
using appropriate space-time and Wick rotations \cite{GrGut}.
In this case we have
\be
\label{NSNS19+}
\langle B p, \eta | e^{-l H} | B (p+8), \eta\rangle_{NSNS} = 16
\left({f_4(e^{-2\pi l}) \over f_2(e^{-2\pi l})}\right)^8 \,,
\ee
\be
\label{NSNS19-}
\langle B p, \eta | e^{-l H} | B (p+8), -\eta\rangle_{NSNS} = 16
\left({f_3(e^{-2\pi l}) \over f_2(e^{-2\pi l})}\right)^8\,.
\ee
The amplitude then becomes in the first case of (\ref{NSNSpp})
% corrected numerical factors in front of integrals and in exponents
\be
\begin{array}{lcl}
\A(p,\eta;p,\eta)_{NSNS} & = & {\displaystyle
{1 \over 2} \int_{0}^{\infty} {dl \over 2}
\left({f_3(e^{-2\pi l}) \over f_1(e^{-2\pi l})}\right)^8} \\[6pt]
& = & {\displaystyle 
{1 \over 8} \int_{0}^{\infty} {dt \over t} t^{-5}
\left({f_3(e^{-\pi t}) \over f_1(e^{-\pi t})}\right)^8 \,,}
\end{array}
\label{c.10}
\ee
where we have made the
substitution $l=1/2t$, and used the modular transformation
properties of the functions $f_i$. We can interpret this as a loop 
calculation in the (unprojected) NS sector of an open string theory. 
By similar manipulations we obtain
\be
\label{NSNSpp-a}
\A(p,\eta;p,-\eta)_{NSNS} = 
{1 \over 8}  \int_{0}^{\infty} {dt \over t} t^{-5}
\left({f_2(e^{-\pi t}) \over f_1(e^{-\pi t})}\right)^8 \,,
\ee
which corresponds to the R sector of an open string. 
Similarly, we have
\be
\label{NSNS19+a}
\A(p,\eta;p+8,\eta)_{NSNS} = 
 2 \int_{0}^{\infty} {dt \over t} t^{-1}
\left({f_2(e^{-\pi t}) \over f_4(e^{-\pi t})}\right)^8 \,,
\ee
and
\be
\label{NSNS19-a}
\A(p,\eta;p+8,-\eta)_{NSNS} = 
 2 \int_{0}^{\infty} {dt \over t} t^{-1}
\left({f_3(e^{-\pi t}) \over f_4(e^{-\pi t})}\right)^8 \,.
\ee
These correspond in the first case to the NS open string sector,
where all eight transverse coordinates are twisted, {\it i.e.}
the fermionic modes are integral and the bosonic modes
are half-integral. The second case is the R open string sector,
where all eight transverse coordinates are twisted. 
The ground state energy of the first open string is
$+1/2$, and that of the second open string is $0$.

In the R-R sector, a similar analysis is possible, and the
results are
\be
\label{RRpp+}
\A(p,\eta;p,\eta)_{RR} = - 
{1 \over 8}  \int_{0}^{\infty} {dt \over t} t^{-5}
\left({f_4(e^{-\pi t}) \over f_1(e^{-\pi t})}\right)^8 \,,
\ee
\be
\label{RRpp-}
\A(p,\eta;p,-\eta)_{RR} = 0\,,
\ee
\be
\label{RR19+}
\A(p,\eta;p+8,\eta)_{RR} = 0\,,
\ee
\be
\label{RR19-}
\A(p,\eta;p+8,-\eta)_{RR} = -
 2 \int_{0}^{\infty} {dt \over t} t^{-1} \,,
\ee
where, as always, $t=1/2l$. These correspond in the first case
to the NS open string sector with $(-1)^F$, and in the last
case to the $R$ open string sector with $(-1)^F$, where all
eight transverse coordinates are twisted. The ground state
energy in the first case is $-1/2$, and in the last case it is $0$.
\medskip

Finally, we should mention that we can use the above techniques
to derive the forces between the various $D$-branes, thereby
recovering the results of \cite{Bach,GrGut,Lif}. For example,
it follows from (\ref{NSNSamp}) and a similar formula
for the R-R sector, that two type II branes whose
dimensions differ by $4$ do not exert any force in either
the NS-NS or R-R sectors onto each other.

\section*{Acknowledgements}

\noindent We would like to thank Joe Polchinski, Cumrun Vafa
and Barton Zwiebach for very helpful discussions. We would also
like to thank Wolfgang Eholzer, Michael Green, Albion Lawrence, 
Zurab Kakushadze and Steve Naculich for helpful comments.
O.B. is supported in part by the NSF
under grants PHY-93-15811 and PHY-92-18167, and M.R.G. is supported by a 
NATO-Fellowship and in part by NSF grant PHY-92-18167.

\end{document}